\documentclass[aps,prl,superscriptaddress,twocolumn,10pt]{revtex4-1}
\usepackage{graphicx}
\usepackage{amssymb}

\newcommand{\sign}{\mathop{\rm sign}\nolimits}

\newcommand{\kF}{k_{\text{F}}}
\usepackage{color}

\usepackage{amsmath}
\newcommand{\angstrom}{\text{\r{A}}}

\usepackage[normalem]{ulem}

\begin{document}

\title{Luttinger liquid behavior of one-dimensional $^3$He}
\author{G. E. Astrakharchik}
\author{J. Boronat}
\affiliation{Departament de F\'{\i}sica i Enginyeria Nuclear, Campus Nord B4-B5, Universitat Polit\`ecnica de Catalunya, E-08034 Barcelona, Spain}
\date{30 December 2014}

\begin{abstract}
The ground-state properties of one-dimensional $^3$He are studied using quantum Monte Carlo methods.
The equation of state is calculated in a wide range of physically relevant densities and is well interoplated by a power-series fit.
The Luttinger liquid theory is found to describe the long-range properties of the correlation functions.
The density dependence of the Luttinger parameter is explicitly found and interestingly it shows a non-monotonic behavior.
Depending on the density, the static structure factor can be a smooth function of the momentum or might contain a peak of a finite or infinite height.
Although no phase transitions are present in the system, we identify a number of physically different regimes, including an ideal Fermi gas, a ``Bose-gas'', a ``super-Tonks-Girardeau'' regime, and a ``quasi-crystal''.
The obtained results are applicable  to unpolarized, partially or fully polarized $^3$He.
\end{abstract}

\maketitle

The experimental realization and study of one-dimensional (1D) quantum fluids is nowadays an active research area.
Recently, a clear evidence of the 1D character was found in $^3$He~\cite{yager} and $^4$He\cite{Savard11,Taniguchi13} confined in narrow nanapores~\cite{substrate}.
One-dimensional Luttinger liquid behavior was also observed in mass flux inside solid $^4$He~\cite{Vekhov12,Vekhov14}.
The Fermi statistics of $^3$He allowed for using nuclear magnetic resonance (NMR) to study the dependence of spin relaxation time with the angular frequency, obtaining a dependence $\omega^{1/2}$ characteristic of 1D diffusion.
It was claimed~\cite{yager} that NMR measures would provide access to the experimental determination of the Luttinger parameter.
Nevertheless, the experiment was carried out at too high temperature (1.7 K) to achieve quantum degeneracy.
There are additional evidences of the 1D behavior of $^3$He by heat capacity measurements~\cite{wada,taniguchi}.
These experimental achievements open the real possibility of obtaining a stable Fermi Luttinger liquid, with the relevant advantage of existing at any density since 1D $^3$He does not sustain a many-body self-bound state.
In addition, and from a theoretical point of view, the properties of $^3$He can be accurately determined because the helium interaction potential is known with high precision~\cite{Aziz87}.


In this paper, we present a quantum Monte Carlo study of one-dimensional $^3$He in the limit of zero temperature, described by the Hamiltonian
\begin{eqnarray}
\hat H
= -\frac{\hbar^2}{2m}\sum\limits_{i=1}^N \Delta_i
+
\sum\limits_{i<j} V(|x_i-x_j|),
\label{H}
\end{eqnarray}
where $m=3.016$u is the $^3$He mass and $x_i, i=\overline{1,N}$ are the positions of each one of the $N$ atoms.
We take the two-body interaction potential $V(r)$ in Aziz~II form~\cite{Aziz87}.
The system is simulated by imposing periodic boundary conditions (p.b.c.) on a box of size $L$.
In the thermodynamic limit, the properties are governed by a single parameter, namely the linear density $\rho = N/L$.

We use the diffusion Monte Carlo (DMC) method to study the properties at $T=0$\cite{DMC}. The DMC algorithm solves the Schr\"odinger equation in imaginary time.
The variance of the results is greatly reduced by introducing an importance sampling based on use of a guiding wave function, which we choose in a pair-product form
\begin{eqnarray}
\psi_\text{T}(x_1, ..., x_N) = \prod\limits_{i<j}^N f_\text{2}(x_i-x_j) \, \sign(x_i-x_j),
\label{wf}
\end{eqnarray}
where $f_{2}(x)$ is the solution of the two-body scattering problem matched for $x>R_{par}$ with the phononic asymptotics $|\sin(\pi x/L)|^{1/K_{par}}$. Parameters $R_{par}$ and $K_{par}$ are optimized by minimizing the variational energy.
The sign function in Eq.~(\ref{wf}) ensures that the wave function changes its sign when any two fermionic atoms are exchanged.
While the guiding wave function based on the two-body solution is known to produce poor energy in 3D, we find that in 1D choice~(\ref{wf}) works very well.
A possible reason is that the three-body collisions are greatly suppressed in reduced geometry.
Furthermore, wave function~(\ref{wf}) becomes exact in the $\rho\to 0$ limit.
Indeed, when the interparticle distance is large, the potential energy is negligible and the system
becomes equivalent to an ideal Fermi gas.
The ground-state wave function corresponds then to a Slater determinant of plane waves, which in 1D can be recasted in a Vandermonde form.
Its explicit evaluation results in expression~(\ref{wf}), with $f_{2}(x) = \sin(\pi x/L)$.
Since the 1D nodal surface is exact the fixed-node approximation gives the energy exactly and thus the sign problem disappears.
According to the Girardeau mapping\cite{Girardeau60}, for 1D interaction potentials with hard-core part, there is a simple relation between fermionic $\psi_\text{F}$ and bosonic $\psi_\text{B}$ wave functions, $\psi_\text{B} = |\psi_\text{F}|$.
This implies that the energy and the diagonal properties (density profile, pair distribution function, etc) in 1D helium are independent of the statistics and polarization.
The crucial difference comes not from the statistics but rather from the atomic mass which is lighter in $^3$He (fermion) than in $^4$He (boson).
In particular, the lighter mass of $^3$He leads to the absence of a liquid phase, which is instead present in $^4$He~\cite{boro_1d4he}.
It is worth noticing that non-local properties (one-body density matrix, momentum distribution, etc) still depend on quantum statistics.
All Monte Carlo results reported here are obtained for $N=100$ particles. In the case of the energy finite-size correction is taken into account.

\begin{figure}
\includegraphics[width=0.45\textwidth,angle=0]{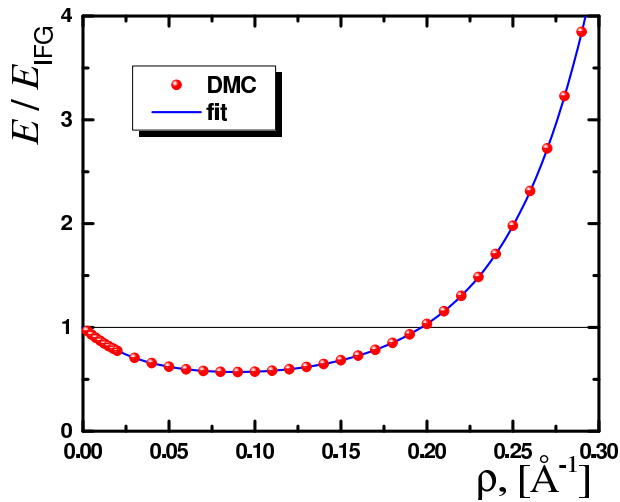}
\caption{(Color online) Ground-state energy $E$ in units of the energy of the ideal Fermi gas $E_{\text{IFG}}$ 
as a function of the linear density $\rho$. Symbols - DMC data, line - fit~(\ref{fit}).
Statistical errors are smaller than the symbol size.}
\label{fig:E}
\end{figure}

The ground-state energy, obtained in DMC calculations, is exact (apart from some controllable statistical error).
In the low density limit, the system behaves as an ideal Fermi gas (IFG) and the energy per particle has a quadratic dependence with the density $\rho$
$\frac{E_{\text{IFG}}}{N} =
\frac{E_\text{F}}{3},
$
where $E_\text{F} = \pi^2\hbar^2\rho^2/2m$ is the Fermi energy.
We find that the ideal Fermi gas limit realizes at densities $\rho\lesssim 0.01$\r{A}$^{-1}$ (see Fig.~\ref{fig:E}).
For larger densities, the attractive long-range part of the He-He interaction contributes significantly to the potential energy.
This ``softening'' of the energy increases up to a density $\rho\approx 0.1$\r{A}$^{-1}$.
For larger densities, the short-range repulsive part makes the system more rigid.
At the density $\rho\approx 0.2$\r{A}$^{-1}$ the energy per particle becomes larger than that of an ideal Fermi gas and the energy
diverges quickly as the density is further increased.
The core size $\sigma=2.963$\r{A} of the Aziz~II potential imposes the maximal density $\rho_{max} = 1/\sigma = 0.3375$\r{A}$^{-1}$, up to which the system can be compressed.
At this density, the excluded hard-core volume fills all available space, resulting in the formation of a true crystal with a diverging energy.
The Aziz II potential cannot be used at such extreme densities.
Nevertheless, we will show that at smaller densities, where the interaction potential is still valid, the system forms a quasi-crystal.

The DMC energy is well interpolated in the range $0<\rho<0.3$\AA$^{-1}$ with a polynomial series
\begin{eqnarray}
\frac{E}{N} = \frac{\pi^2 \hbar^2 \rho^2}{6m}\left(1+\sum_{i=1}^6 C_i \rho^i \right)
\label{fit}
\label{EoS}
\end{eqnarray}
with coefficients
$C_1 = -14.91(5)$\AA,
$C_2 = 219.6(7)$\AA$^2$,
$C_3 = -1915(4)$\AA$^3$,
$C_4 = 10533(10) $\AA$^4$,
$C_5 = -31160(50)$\AA$^5$, and
$C_6 = 41725(100)$\AA$^6$
(figures in parentheses are the errors on the last digits).
While compared to an ideal Fermi gas, $^3$He might be both softer or more rigid, the energy per particle always remains a monotonously increasing function of the density
(note that Fig.~\ref{fig:E} reports the energy divided by the energy of an ideal Fermi gas).
This should be contrasted with $^4$He, where its heavier mass leads to the existence of a minimum in the energy, physically reflecting the formation of a liquid (many-body self-bound) state.
Contrarily, one-dimensional $^3$He always stays in a gas state.

\begin{figure}
\includegraphics[width=0.45\textwidth,angle=0]{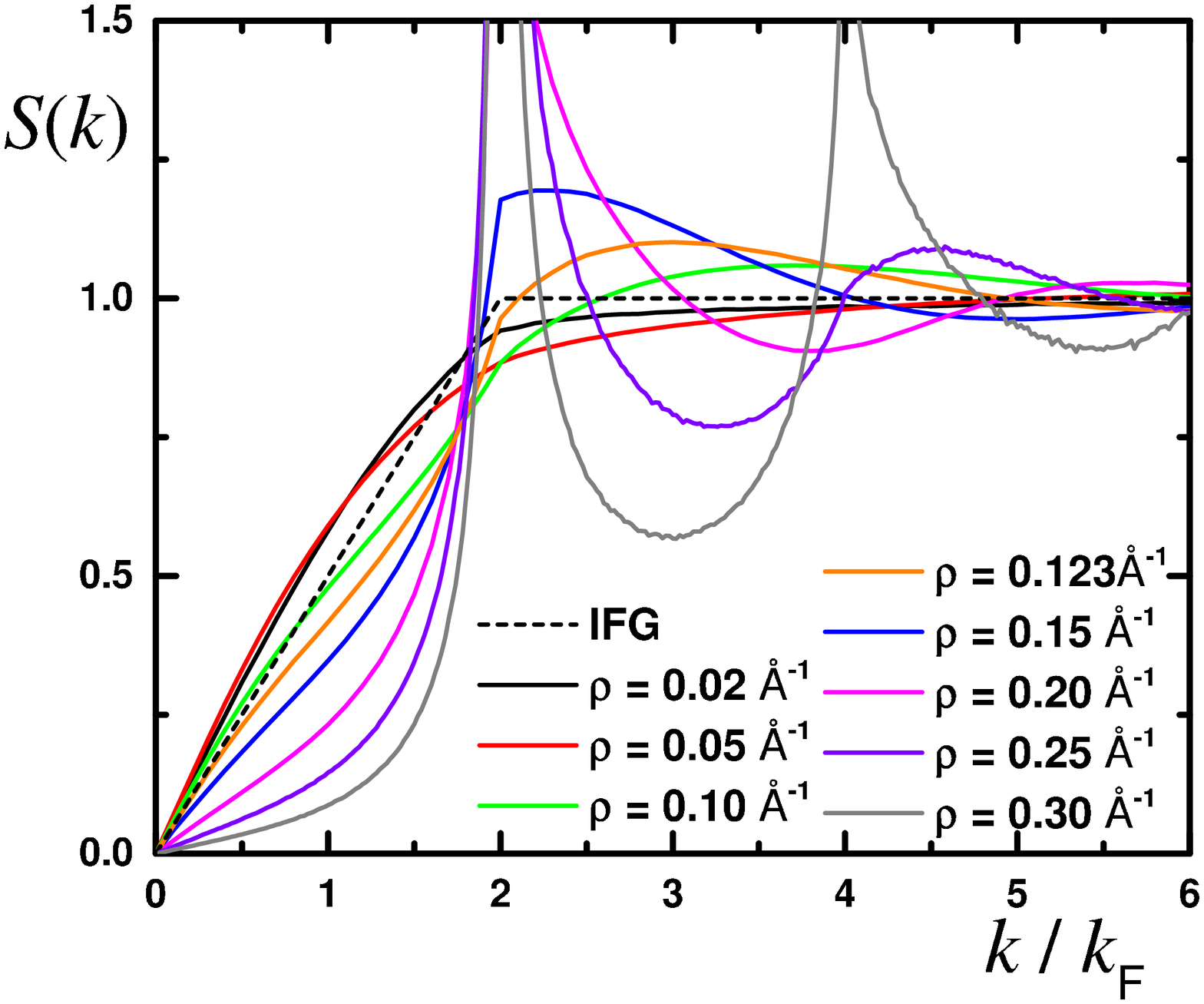}
\caption{(Color online) Static structure factor $S(k)$ in units of the Fermi momentum $\kF=\pi\rho$. Dashed line -- ideal Fermi gas.}
\label{fig:Sk}
\end{figure}

In order to study the structural properties, we calculate the static structure factor $S(k)$ using the technique of pure estimators~\cite{pure}.
The results obtained for a wide range of densities are shown in Fig.~\ref{fig:Sk}.
In a dilute system, the static structure factor approaches the one of an ideal Fermi gas.
In this case, the low-momentum linear behavior
\begin{eqnarray}
S(k) = \hbar |k|/2mc, \qquad k\to 0
\label{Sk:phonons}
\end{eqnarray}
continues up to $|k|=2\kF$ with $\kF = \pi\rho$ the Fermi momentum.
The slope is fixed by the speed of sound which for the ideal Fermi gas is $c = v_\text{F} = \hbar \kF/m$.
For any larger momenta, $|k|>2\kF$, the static structure factor is equal to the large-momentum asymptotic value $S(k)=1$.
The first derivative is discontinuous at $|k|=2\kF$ corresponding to an {\it umklapp} process in which an atom is flipped from one side of the Fermi surface to the other.
The kink at $|k| = 2k_\text{F}$ disappears as the density is increased and the attractive part of the interaction potential becomes relevant.
In this regime, $S(k)$ is a smooth monotonous function with the typical shape found in weakly interacting Bose gases (see, for instance, the Lieb-Liniger model~\cite{Sk}).
The Gross-Pitaevskii prediction for the speed of sound in a Bose gas, $c = \sqrt{g_\text{1D}\rho/m}$ with $g_\text{1D}$ the coupling constant\cite{PS}, leads to a linear slope in $S(k)$ which is larger than that of an ideal Fermi (or Tonks-Girardeau) gas.
Although the Gross-Pitaevskii regime is never reached in helium system, we still see an increase in the slope of $S(k)$ in the regime of equivalence to a Bose gas system ($\rho \lesssim 0.05$\AA).
On the contrary,  at larger density the linear slope starts to decrease, reflecting the growth of the contribution coming from the hard-core part of the interaction potential.
Eventually, a kink at $k=2\kF$ is formed which, as the density is increased further, gets transformed to a sharp peak.
A diverging peak in $S(k)$ manifests the formation of a quasi-crystal.
In fact, the wave number $k=2\kF$ corresponds to the width of the first Brillouin zone $2k_\text{F}=2\pi/a$ in a crystal with the lattice spacing  $a=\rho^{-1}$.
It should be noted that the magnetic structure factor trivially vanishes $S_\text{M}(k) = S_{\uparrow\uparrow}(k) - S_{\uparrow\downarrow}(k) = 0$, as according to Girardeau's mapping $S_{\uparrow\uparrow}(k) = S_{\uparrow\downarrow}(k) = S(k)$.

In order to define the boundaries between ``Bose-gas'', ``super-Tonks-Girardeau'' and ``quasi-crystal'' regimes we make use of the Luttinger liquid theory\cite{Haldane, Cazalilla04, LuttingerLiquid}.
This effective theory applies to systems with gapless excitation spectrum $E_k = \hbar |k| c$ and predicts the large-distance (or small-momenta) behavior of the correlation functions.
The results are universal in terms of the Luttinger parameter which, in a homogeneous system, is directly related to the
Fermi velocity $v_\text{F}$ and the speed of sound $c$ as $K = v_\text{F}/c$.
The Fermi velocity $v_\text{F}$ is completely fixed by the density while the speed of sound $c$ takes into account the many-body interactions between particles.
The description in terms of a Luttinger liquid is very broad and applies to a large number of one-dimensional systems, but a microscopic calculation is always necessary to relate the Luttinger parameter $K$ to the linear density $\rho$.

\begin{figure}
\includegraphics[width=0.45\textwidth,angle=0]{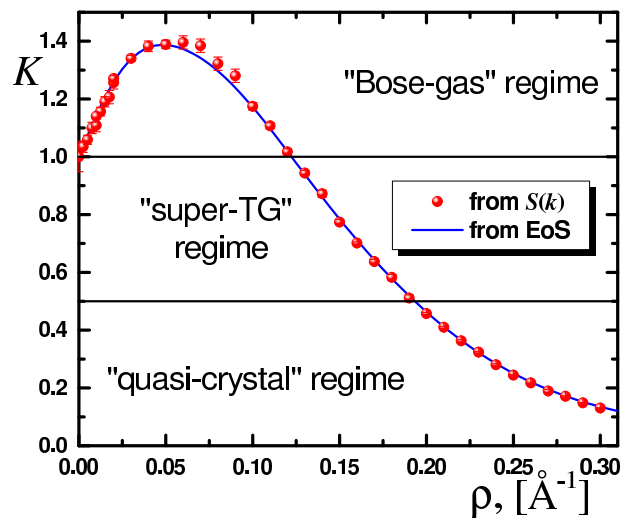}
\caption{(Color online) Luttinger parameter $K$ as a function of the linear density $\rho$. Symbols - speed of sound is extracted from the phononic part of the static structure factor, line - speed of sound is extracted from the compressibility using fit.~(\ref{fit}) to the equation of state.}
\label{fig:K}
\end{figure}

The dependence of $K$ on $\rho$ is reported in Fig.~\ref{fig:K} and constitutes the main result of this paper.
We use two alternative approaches to find the speed of sound $c$ and to establish the dependence of $K$ on $\rho$:
(i) by differentiating the equation of state~(\ref{fit}) we obtain the chemical potential
$\mu = \partial E/\partial N$ and, from that, the speed of sound $c$,
$mc^2 = \rho \partial \mu/\partial \rho$
(ii) from the linear behavior of $S(k)$ with $k$ at low momentum (\ref{Sk:phonons}).
The fact that two different approaches are in agreement demonstrates the high precision of the calculations and the internal consistency of  our approach.
In the low-density limit, the value of the Luttinger parameter, $K=1$, corresponds to that of an ideal Fermi gas where $c=v_\text{F}$.
By increasing the density, the Luttinger parameter becomes larger, $K>1$, corresponding to a Fermi system with dominant attraction and similarly to an interacting Bose gas (Lieb-Liniger model)\cite{LiebLiniger63,Cazalilla04}.
The non-monotonous behavior with the density makes the Luttinger parameter to return at $\rho\approx 0.123$\r{A}$^{-1}$ to $K=1$, which can be associated to the Tonks-Girardeau regime for bosons.
For larger density, $K$ becomes smaller than unity, $K<1$.
This regime corresponds to a Fermi system with a dominant repulsion and is similar in many aspects to the super-Tonks-Girardeau regime for bosons~\cite{sTG}.
The ``quasi-crystal'' regime is reached for $K<1/2$, at $\rho \gtrsim 0.19\angstrom^{-1}$, and is characterized by a diverging peak in the static structure factor at the momentum corresponding to the inverse lattice spacing.
The reentrant behavior, when two different densities correspond to the same value of the Luttinger parameter, is a peculiar feature of the helium interaction and of the gas nature of $^3$He, and was not observed in purely repulsive systems, such as $\delta$-interacting bosons\cite{Cazalilla04}, hard rods\cite{hardrods}, dipoles\cite{K:dipoles,dipoles1D} or Coulomb charges\cite{coulomb1D}.
This rich behavior in helium is caused by a competition between repulsive hard core and an attractive van der Waals tail in the interaction potential.

Within the Luttinger liquid theory, the pair distribution function $g_2(x)$ can be expanded at large distances in a series of oscillating terms with a power-law envelope\cite{Haldane}
\begin{eqnarray}
\frac{g_2(x)}{\rho^2} = 1 - \frac{K}{2(\kF x)^2} +
\sum\limits_{l=1}^{\infty} A_l \frac{\cos (2l\kF x)}{|\kF x|^{2l^2K}} \ .
\label{g2}
\end{eqnarray}
The terms on the r.h.s. of Eq. (\ref{g2}) describe
(i) constant value of uncorrelated particles,
(ii) $1/x^2$ decay due to density fluctuations, with the amplitude fully fixed by $K$, and
(iii) oscillations with a power-law envelope.
The exponents in the power-law decay are fully fixed by $K$, while the amplitudes $A_l$ cannot be established within the Luttinger liquid approach.
Such oscillations might cause divergencies at the multiples of the lattice momentum $k = 2l\kF = l(2\pi/a)$.
The height of the $l$-th peak
\begin{eqnarray}
S(k=2l\kF) = A_l N^{1-2l^2K}
\label{Eq:Sk:peak}
\end{eqnarray}
diverges when $K<1/(2l^2)$.
In particular, the first peak diverges when $K<1/2$, i.e., for densities $\rho \gtrsim 0.2\angstrom^{-1}$.
We call this regime a ``quasi-crystal''.
There is a number of differences between ``quasi'' and ``true'' crystals.
A three-dimensional crystal possesses a diagonal long-range order as the density oscillations remain in phase for large distances.
Instead, in one dimension the order is lost in a power-law way.
The height of the peak is divergent in both cases, but in a true crystal the Bragg peak grows linearly with the number of particles $S(k_{peak}) \propto N$, while in a quasi-crystal the exponent is smaller than unity, as can be seen from Eq.~(\ref{Eq:Sk:peak}).
A true crystal is characterized by a number of diverging peaks, while in one dimension the number of peaks depend on the value of $K$. There are no diverging peaks for $K>1/2$, one diverging peak for $1/8<K<1/2$, two for $1/18<K<1/8$ and so on. Only when asymptotically $K\to 0$ (or $\rho\to\rho_{max}$) the true crystal is recovered.

\begin{figure}
\includegraphics[width=0.45\textwidth,angle=0]{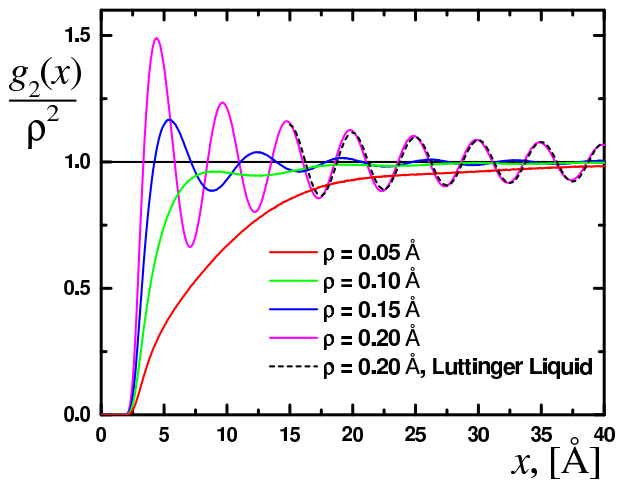}
\caption{(Color online) Pair distribution function $g_2(x)$ for $N=20$ atoms at different densities (high of the peak decreases from the highest to the lowest density). Short-dashed line, Luttinger liquid asymptotics, Eq.~(\ref{g2})\cite{Note1}.
}\label{fig:g2}
\end{figure}

Knowledge of the Luttinger parameter opens up the possibility to use effective Hamiltonian theories to predict how the helium system will behave in the presence of disorder or a periodic lattice~\cite{Cazalilla11}.
A single impurity is irrelevant for $K>1$ and leads to pinning for $K<1$~\cite{single impurity},
which might be absent in larger pores~\cite{DelMaestro11}.
The effect is stronger for a continuous density, where the disorder induces
superfluid-to-insulator transition at the critical value of $K=3/2$~\cite{Giamarchi,Prokofiev}.
In this case, 1D $^3$He stays always in the same phase of incompressible insulator
(Bose glass~\cite{Fisher89}).
In the case of 1D periodic lattice with filling fraction $1/p$, the transition to the Mott insulator state happens for $K=2/p^2$~\cite{Cazalilla11}.

The pair distribution function is shown in Fig.~\ref{fig:g2} for a number of characteristic densities.
The short-range behavior is dominated by the two-body physics, where the repulsive hard core make it impossible for the helium atoms to approach each other for distances smaller than $x\lesssim 2\angstrom$.
The long-range behavior has oscillations at multiples of $2k_\text{F}$ modulated with a power-law envelope~(\ref{g2})\cite{Boninsegni13}.

The form factor touches zero for $k=2k_\text{F}$ in a power-law way, $S(\omega, 2k_\text{F})\sim\omega^{2(K-1)}$\cite{Castro}.
Contrary to three dimensions, where the excitation spectrum has a roton minimum and helium is superfluid according to Landau argument, in one dimension the excitation spectrum always touches zero.
A drag of impurity through the system always leads to dissipation of energy\cite{BECimpurity}
and phase slips induce~\cite{Eggel11} decay, making the gas normal even at zero temperature. Moreover, in one dimension quantum fluctuations destroy the diagonal long-range order\cite{Mermin} and no phase transition is possible at finite temperature\cite{LandauLifshitz}. The system always stays in the same phase which is not superfluid, is not Bose-condensed and does not form a true crystal.

To summarize, we have studied the ground-state properties of one-dimensional $^3$He
by means of the diffusion Monte Carlo method.
The helium-helium interaction potential is known very precisely permitting us to make
quantitative predictions for the energy and the correlation functions, which can be
experimentally measured.
The equation of state is obtained for a wide range of densities with the fit given by
Eq.~(\ref{fit}).  The structural properties are addressed by studying the static
structure factor $S(k)$ and the pair distribution function $g_2(r)$.
It is worth noticing that 1D geometry makes that all these results are the same in polarized and unpolarized $^3$He.
We show that the long-range properties of the correlation functions are well reproduced by the Luttinger liquid theory.
We extract the Luttinger parameter $K$  (i) from the compressibility by using the fit to the equation of state
(ii) from the linear slope of $S(k)$ for small momenta.
Both methods agree within the error bars proving a high quality of the results and the internal consistency of the method.
The obtained relation between the Luttinger parameter $K$ and the linear density $\rho$ can be used to predict the long-range asymptotics of stationary correlation functions (one-body density matrix, momentum distribution, etc) as well as dynamic quantities (dynamic form factor $S(k,\omega)$ close to $k=2k_\text{F}$, etc).
The non-monotonous dependence of $K$ on the density was not previously observed
in fully repulsive systems and is a special feature of $^3$He.
Remarkably, we found that a particular value of the Luttinger parameter might be attained at two different densities.
Although no phase transitions are present, we identify the following physically different regimes:
(i) $\rho\lesssim 0.01\angstrom^{-1}$, ideal Fermi gas regime
(ii) $0.01\angstrom^{-1}<\rho<0.123\angstrom^{-1}$, ``Bose-gas'' regime corresponding to a Fermi system
with a dominant attraction and long-range behavior of the correlation functions
similar to that of a repulsive Bose gas,
(iii) $0.123\angstrom^{-1}<\rho<0.19\angstrom^{-1}$, ``super-Tonks-Girardeau'' (sTG) regime corresponding to a Fermi system with dominant repulsion, characterized by the formation of a peak in $S(k)$, similar to a Bose system in sTG regime, and
(iv) $\rho>0.19\angstrom^{-1}$, ``quasi-crystal'' regime, characterized by a
 diverging peak in $S(k)$ at $k=2k_\text{F}$.
We expect that our results can provide quantitatively precise predictions on the Luttinger liquid behavior of 1D $^3$He which can be verified in future experiments.

We acknowledge financial support from the DGI (Spain) Grant No. FIS2011-25275 and the Generalitat de Catalunya Grant No. 2009SGR-1003. GEA acknowledges a fellowship by MEC (Spain) through the Ram\'on y Cajal program.
Part of the numerical simulations was carried out at Barcelona Supercomputing Center (The Spanish National Supercomputing Center -- Centro Nacional de Supercomputaci\'on).

\end{document}